**Cognitive control as a multivariate optimization problem**

Harrison Ritz*, Xiamin Leng, & Amitai Shenhav

*Cognitive, Linguistic, and Psychological Sciences, Carney Institute for Brain Science, Brown University, Providence, RI*

* Corresponding author (harrison.ritz@gmail.com)

**Abstract**

A hallmark of adaptation in humans and other animals is our ability to control how we think and behave across different settings. Research has characterized the various forms cognitive control can take – including enhancement of goal-relevant information, suppression of goal-irrelevant information, and overall inhibition of potential responses – and has identified computations and neural circuits that underpin this multitude of control types. Studies have also identified a wide range of situations that elicit adjustments in control allocation (e.g., those eliciting signals indicating an error or increased processing conflict), but the rules governing when a given situation will give rise to a given control adjustment remain poorly understood. Significant progress has recently been made on this front by casting the allocation of control as a decision-making problem. This approach has developed unifying and normative models that prescribe when and how a change in incentives and task demands will result in changes in a given form of control. Despite their successes, these models, and the experiments that have been developed to test them, have yet to face their greatest challenge: deciding how to select among the multiplicity of configurations that control can take at any given time. Here, we will lay out the complexities of the inverse problem inherent to cognitive control allocation, and their close parallels to inverse problems within *motor* control (e.g., choosing between redundant limb movements). We discuss existing solutions to motor control's inverse problems drawn from optimal control theory, which have proposed that effort costs act to regularize actions and transform motor planning into a well-posed problem. These same principles may help shed light on how our brains optimize over complex control configuration, while providing a new normative perspective on the origins of mental effort.

Keywords: Cognitive control; Motor control; Decision making; Inverse problem

*"There are many paths up the mountain, but the view from the top is always the same"*
- Chinese Proverb

Over the past half-century, our understanding of the human brain's capacity for cognitive control has grown tremendously (Abrahamse et al., 2016; Botvinick and Cohen, 2014; Fortenbaugh et al., 2017; Friedman and Robbins, 2021; Koch et al., 2018; Menon and D'Esposito, 2021; von Bastian et al., 2020; Westbrook and Braver, 2015). The field has developed consistent ways of defining and operationalizing control, such as in terms of its functions and what distinguishes different degrees of automaticity (Cohen et al., 1992; Posner and Snyder, 1975; Shiffrin and Schneider, 1977). It has developed consistent methods for eliciting control and measuring the extent to which control is engaged by a given task (Danielmeier and Ullsperger, 2011; Egner, 2007; Gonthier et al., 2016; Koch et al., 2018; von Bastian et al., 2020; Weichart et al., 2020). It has demonstrated how such control engagement varies across individuals (Friedman and Miyake, 2017; von Bastian et al., 2020) and over the lifespan (Braver and Barch, 2002; Luna, 2009). Finally, research in this area has made substantial progress towards mapping the neural circuitry that underpins the execution of different forms of cognitive control (Friedman and Robbins, 2021; Menon and D'Esposito, 2021; Parro et al., 2018; Shenhav et al., 2013). The factors that determine *how* cognitive control is configured have, on the other hand, remained mysterious and heavily debated (Shenhav et al., 2017).

Studies have uncovered reliable antecedents for control adjustments, including the commission of an error (Danielmeier and Ullsperger, 2011; Rabbitt, 1966) or changes in task demands (Gratton et al., 1992; Logan and Zbrodoff, 1979). However, it has been a longstanding goal for the field to develop a comprehensive model of how people use the broader array of information they monitor to configure the broader array of control signals they can deploy. To address this question, models proposed that the problem of determining control allocation can be solved through a general decision-making process that involves weighing the costs and benefits of potential control allocations (Lieder et al., 2018; Shenhav et al., 2013; Verguts et al., 2015; Westbrook and Braver, 2015). These models have already shown promise in accounting for how people adjust *individual* control signals (e.g., how much to adjust attention towards a particular task) based on the incentives and demands of a given task environment (Bustamante et al., 2021; Lieder et al., 2018; Musslick et al., 2015; Verguts et al., 2015). Here, we focus on a different aspect of this problem: how is it that people navigate the *multitude* of solutions that can match the demands of their environment? How can cognitive control scale to configuring the complex information processing we deploy throughout our daily life? What is the relationship of mental effort to the multiplicity of options for configuring control? Building off well-characterized computational models from motor planning, we examine how multiplicity presents a critical challenge to cognitive control configuration, and how algorithmic principles from motor control can help to overcome these challenges and refine our understanding of goal-directed cognition.

# The multiplicity of cognitive control

To study the mechanisms that govern the allocation of cognitive control, researchers have sought to identify reliable predictors of *changes* in control allocation within and across experiments. These triggers for control adjustment have in turn provided insight into signals – such as errors and processing conflict – that the brain could monitor to increase or decrease control. Research has shown that control adjustments induced by these signals, even within the same setting, vary not only in degree but also kind (see Table 1).

*Error-related control adjustments*

In common cognitive control tasks such as the Stroop, Simon, and Eriksen flanker task (Egner, 2007; von Bastian et al., 2020), participants have prepotent biases that often lead to incorrect responses (e.g., responding based on the salient flanking arrows rather than the goal-relevant central arrow). Errors thus serve a useful indicator that the participant was likely under-exerting control and should adjust their control accordingly. The best-studied instantiation of error-related control adjustments manifests in a participant's tendency to respond more slowly and more accurately after an error (Danielmeier and Ullsperger, 2011; Laming, 1979a; Rabbitt, 1966), which can be understood as together reflecting post-error adjustments in caution. Indeed, work has shown that models like the drift diffusion model[1] (DDM; Ratcliff, 1978; Ratcliff and McKoon, 2008; see Figure 1A), post-error slowing and post-error increases in accuracy can be jointly accounted for by an increase in one's *response threshold*, the criterion they set for how much evidence to accumulate about the task stimuli before deciding how to respond (Dutilh et al., 2012; Fischer et al., 2018).

Experiments investigating the neural implementation of these post-error adjustments have found that threshold adjustments are associated with the suppression of motor-related activity (Danielmeier et al., 2011; Fischer et al., 2018). For instance, Danielmeier and colleagues (2011) had participants perform a Simon-like task that required them to respond based on the color of an array of dots that were moving in a direction compatible or incompatible with the correct color response. When participants responded incorrectly, they tended to be slower and more accurate on the following trial. This increased caution was coupled with decreased BOLD activity in motor cortex on that subsequent trial, consistent with the possibility that errors led to controlled adjustments of decision threshold (in this case by putatively lowering the baseline activity to require more evidence before responding).

In addition to changing overall caution, errors can also influence how specific stimuli are processed. Studies have shown that error trials can be followed by selective enhancement of task-relevant (target) processing (Danielmeier et al., 2015, 2011; King et al., 2010; Maier et al., 2011; Steinhauser and Andersen, 2019) and/or suppression of task-irrelevant (distractor) processing (Danielmeier et al., 2015, 2011; Fischer et al., 2018)). For instance, in the same study by Danielmeier and colleagues (2011), errors tended to be followed by increased activity in regions encoding the target stimulus dimension and decreased activity in regions encoding the distractor dimension (see also (Fischer et al., 2018; King et al., 2010)). Thus, whereas post-error slowing effects reflect control over one's decision threshold, such post-error reductions of interference likely reflect a different form of control, one that adjusts the influence of target- and distractor-related information on the evidence that is accumulated before reaching that threshold (target and distraction contribution to the *drift rate* in the DDM).

---

[1] Note that the DDM shares properties with several other evidence accumulation models that enable similar behavioral predictions, and in some cases finer-grained predictions for neural implementation (Bogacz, 2007). We focus on the DDM as a reference point through much of this article because its properties have been closely studied from the theoretical and empirical perspective and it lends itself well to mechanistic hypotheses, but our attributions to this model and its parameters should be seen as potentially generalizable to related models.

*Conflict-related control adjustments*

In addition to error commission, another potential indicator of insufficient control is the presence of processing conflict (Berlyne, 1957; Botvinick et al., 2001), such as when a person feels simultaneously drawn to respond left (e.g., based on target information) and right (e.g., based on a distractor). One of the best-studied forms of conflict-related control adjustment is the *conflict adaptation* or *congruency sequence effect*, which manifests as reduced sensitivity to response (in)congruency after a person has previously performed one or more high-conflict (e.g., incongruent) trials (Egner et al., 2007; Egner and Hirsch, 2005; Funes et al., 2010; Gratton et al., 1992; Jiang and Egner, 2013). These adaptations are analogous to examples of post-error reductions of interference described above, and have the same candidate computational underpinnings in adjustments to the rate of evidence accumulation (Kerns et al., 2004; Musslick et al., 2015, 2019b) These control adjustments have likewise been found to be associated with changes in neural activity in goal-relevant perceptual processing pathways. For example, Egner & Hirsch (2005) showed that participants were less sensitive to Stroop incongruence after higher-conflict trials, and that this was coupled with increased activity in the target-associated cortical areas (fusiform face area for face targets).

Another body of work has shown that conflict can trigger changes to response threshold, particularly within a trial, for instance when selecting between two similarly-valued options (Aron, 2007; Cavanagh et al., 2011; Fontanesi et al., 2019; Frank et al., 2015; Ratcliff and Frank, 2012; Verguts et al., 2011; Wiecki and Frank, 2013). These adjustments have been linked to interactions between dorsal anterior cingulate cortex (dACC) and the subthalamic nucleus (Brittain et al., 2012; Cavanagh et al., 2011; Frank et al., 2015; Schroeder et al., 2002; Wessel et al., 2019). For instance, simultaneous EEG-fMRI has revealed that BOLD in dACC and mediofrontal EEG theta power moderates the relationship between decision conflict and adjustments to response threshold (Frank et al., 2015).

*Incentive-related control adjustments*

In addition to signals like error and conflict that reflect dips in performance, the need for control can also be signaled by the presence of performance-based incentives (e.g., monetary rewards for good performance). Incentives can influence overall performance, for instance often leading participants to perform tasks faster and more accurately across trials (Parro et al., 2018; Yee and Braver, 2018). Incentives can also trigger task-specific adjustments of cognitive control, enhancing the processing of goal-relevant information (Etzel et al., 2016; Krebs et al., 2010; Soutschek et al., 2014) and/or suppressing the processing of distractor information (Padmala and Pessoa, 2011), likely reflecting changes in associated drift rates similar to error-related adjustments discussed above (cf. Ritz & Shenhav, 2021, discussed further below). Also similar to error-related findings, there is evidence that incentive-related control adjustments are mediated by changes in processing within stimulus-selective circuits (Esterman et al., 2017; Etzel et al., 2016; Hall-McMaster et al., 2019; Padmala and Pessoa, 2011; Soutschek et al., 2015). For example, Padmala and Pessoa (2011) used a Stroop task to show that participants are less sensitive to distractor information when under performance-contingent rewards. They found that this distractor inhibition was mediated by reduced activation in cortical areas sensitive to the distracting stimuli (visual word form area for text distractors).

Performance incentives have been shown to influence not only how well one performs on a given trial but also how *consistently* they perform within and across trials. When performing sustained attention tasks that require participants to repeat the same response on most trials (e.g., frequent go trials) but respond differently on rare occurrences of a different trial type (e.g., infrequent no-go trials), attentional lapses can manifest as increased variability in response times across trials (Fortenbaugh et al., 2017). When performance is incentivized, participants demonstrate both higher accuracy and lower response time variability (Esterman et al., 2017, 2016, 2014). These performance improvements can be accounted for by assuming that incentives influence control over how noisily evidence is accumulated within each trial (e.g., due to mind-wandering; (Ritz et al., 2020); Manohar et al., 2015). Neuroimaging studies suggest that enacting the control required to achieve more consistent (less variable) performance is associated with increases in both sustained and evoked responses in domain-general attentional networks and stimulus-specific regions (Esterman et al., 2017).

|  |  | **Behavior** | **Cognitive process (DDM)** | **Neuroscience** |
|---|---|---|---|---|
| **Errors** | | **RT ↑** | **Threshold ↑** | **Motor cortex activation ↓** |
| | | (Danielmeier et al., 2011; Debener et al., 2005; Gehring and Fencsik, 2001; Jentzsch and Dudschig, 2009; King et al., 2010; Rabbitt, 1966) | (Dutilh et al., 2012; Fischer et al., 2018) | (Danielmeier et al., 2011; King et al., 2010) |
| | | **Error Rate ↓** | | |
| | | (Danielmeier et al., 2011; Laming, 1979b, 1968; Maier et al., 2011; Marco-Pallarés et al., 2008) | | |
| | | **Interference ↓** | **Distractor Drift Rate ↓** | **Target-related activation ↑** |
| | | (King et al., 2010; Maier et al., 2011; Ridderinkhof, 2002; Steinhauser and Andersen, 2019) | (Fischer et al., 2018) | (Danielmeier et al., 2011; King et al., 2010; Maier et al., 2011; Steinhauser and Andersen, 2019) |
| | | | | **Distractor-related activation ↓** |
| | | | | (Danielmeier et al., 2011; Fischer et al., 2018; King et al., 2010) |
| **Conflict** | | **RT ↑** | **Threshold ↑** | **STN activation ↑** |
| | | (Herz et al., 2016; Verguts et al., 2011) | (Fontanesi et al., 2019; Herz et al., 2016) | (Aron, 2007; Cavanagh et al., 2011; Frank et al., 2015; Ratcliff and Frank, 2012; Wiecki and Frank, 2013) |
| | | **Interference ↓** | **Distractor Drift Rate ↓** | **Target-related activation ↑** |

| | | | |
|---|---|---|---|
| | (Braem et al., 2012; Danielmeier et al., 2011; Funes et al., 2010; Gratton et al., 1992; Kerns, 2006; Kerns et al., 2004; Ullsperger et al., 2005) | (Ritz and Shenhav, 2021) | (Egner et al., 2007; Egner and Hirsch, 2005) |
| **Incentives** | **RT ↓  Accuracy ↑** | **Threshold ↑** | |
| | (Chiew and Braver, 2016; Fröber and Dreisbach, 2014; Frömer et al., 2021; Ličen et al., 2016; Soutschek et al., 2014; Yee et al., 2016) | (Dix and Li, 2020; Leng et al., 2020; Thurm et al., 2018) | |
| | | **Threshold ↓** | |
| | | (Leng et al., 2020) | |
| | **Target effect ↑** | **Drift rate ↑** | **Target-related activation ↑** |
| | (Adkins and Lee, 2021; Krebs et al., 2010) | (Dix and Li, 2020; Jang et al., 2021; Leng et al., 2020) | (Etzel et al., 2016; Grahek et al., 2021; Soutschek et al., 2015) |
| | **Distractor effect ↓** | **Target Drift Rate ↑** | **Distractor-related activation ↓** |
| | (Chiew and Braver, 2016; Padmala and Pessoa, 2011; Soutschek et al., 2014) | (Ritz and Shenhav, 2021) | (Padmala and Pessoa, 2011) |
| | **RT variability ↓** | **Accumulation noise ↓** | **Sustained task-relevant activation ↑** |
| | (Esterman et al., 2016, 2014) | (Manohar et al., 2015; Ritz et al., 2020) | (Esterman et al., 2017) |

**Table 1. Multiplicity of control adaptations in response to errors, conflict, and incentives.**

## *Multidimensional configuration of cognitive control*

Previous research has uncovered a multiplicity of adjustments that occur in response to changes in the demands or incentives for control. Importantly, they show that a monitored signal[2] (e.g., an error) can produce several different control adjustments, and that a control adjustment (e.g., increased caution) can be elicited by several different monitored signals. Rather than a strict one-to-one relationship between monitored signals and control adjustments, this diversity suggests that participants make simultaneous decisions across multiple control effectors.

This control multiplicity is evident in studies of post-error adjustments discussed above (Danielmeier and Ullsperger, 2011), in which errors can result in both increased caution (i.e., more conservative response thresholds) and a change in attentional focus to favor target over distractor information (putatively underpinned by adjustments in drift rate). Experiments have found that both adjustments appear to occur simultaneously (Danielmeier et al., 2015, 2011; Fischer et al., 2018; King et al., 2010), reflecting a multi-faceted response to the error event.

---

[2] We will use the term 'monitored signal' to refer to signals that act as inputs to decisions about control allocation. In contrast, we use 'control signals' to refer to the control that is allocated as a result of this decision process (analogous to 'motor commands') (Shenhav et al., 2013).

In a recent experiment, we showed that people can also exert independent control over their processing of targets and distractors (Ritz and Shenhav, 2021)). Like Danielmeier and colleagues (2011), we had participants perform a random dot kinematogram that required responding to dot color while ignoring dot motion. Across trials, we parametrically varied both the target coherence (how easily the correct color could be identified), and distractor interference (how coherently dots were moving in the same or opposite direction as the target response). We found that participants exerted control over their processing of both target and distractor information, but that they did so independently and differentially depending on the relevant task demands. Under performance incentives, participants preferentially enhanced their target sensitivity, whereas after high-conflict trials, participants preferentially suppressed their distractor sensitivity (and, to a lesser extent, also enhanced target sensitivity). A similar pattern has been observed at the neural level while participants perform a Stroop task (Soutschek et al., 2015). Whereas performance incentives preferentially enhanced sensitivity in target-related areas (visual word form area for text targets), conflict expectations preferentially suppressed sensitivity in distractor-related areas (fusiform face area for face distractors). These findings demonstrate that control can be flexibly reconfigured across multiple independent control signals to address relevant incentives and task demands.

There is also evidence that different people prioritize different control strategies within the same setting. For instance, Boksem and colleagues (2006) had participants perform the Simon task over an extended experimental session, and observed performance fatigue in the form of slower and less accurate responding over time. Towards the end of the session, the experimenters introduced monetary incentives and found that this counteracted the effects of fatigue, but did so heterogeneously across the group. When making an error during this incentivized period, some participants responded by focusing more on responding *quickly* while others focused on responding *accurately*. The engagement of these differential control strategies was associated with changes in distinct event-related potentials (error-related negativity vs. contingent negative variation). Similar variability in reliance on different control strategies has been seen across the lifespan (Braver and Barch, 2002; Fortenbaugh et al., 2015; Luna, 2009; Ritz et al., 2020) and between clinical and healthy populations (Casey et al., 2007; Grahek et al., 2019; Lesh et al., 2011).

Collectively, previous research suggests that there is a many-to-many mapping between the information that participants monitor related to task demands, performance, and incentives, and the multitude of control signals that participants can deploy. Recent theoretical models have explained this heterogeneity in terms of the flexible deployment of control, proposing that there is an intervening *decision* process that integrates monitored information, determining which strategies to engage, and to what extent, based on the current situation (Lieder et al., 2018; Shenhav et al., 2013; Verguts et al., 2015).

# Selection and configuration of multivariate control

Casting control allocation as a decision process provides a path toward addressing how people integrate information from their environment to select the optimal control allocation. This process of *optimization* entails finding the best solution for an objective function and set of constraints. Objective functions define the costs and benefits of different solutions, whereas soft constraints (e.g., costs) and hard constraints (e.g., boundary conditions) limit the *space* of possible solutions. Optimization has long played a central and productive role in building computational accounts of multivariate planning in the domain of *motor* control (Flash and Hogan, 1985; Shadmehr and Ahmed, 2020; Todorov and Jordan, 2002; Uno et al., 1989; Wolpert and Landy, 2012), suggesting that this research into how the brain coordinates actions may thus offer general principles for how the brain coordinates cognition.

The starting point for solving any optimization problem is identifying the objective function. Researchers in decision making and motor control have suggested that participants maximize the amount of reward harvested per unit time (*reward rate*; (Harris and Wolpert, 2006; Manohar et al., 2015; Niv et al., 2007; Shadmehr et al., 2010). Studies have found that people's motor actions are sensitive to incentives, with faster and/or more accurate movement during periods when they can earn more rewards (Adkins et al., 2021; Codol et al., 2021, 2020; Manohar et al., 2019, 2017, 2015; Pekny et al., 2015; Sukumar et al., 2021; Trommershäuser et al., 2003a, 2003b; Yoon et al., 2020). For example, participants will saccade toward a target location more quickly and more precisely on trials that are worth more money (Manohar et al., 2019, 2017, 2015). Responding faster and more accurately breaks the traditional speed-accuracy trade-off (Bogacz et al., 2006; Manohar et al., 2015), and is thought to reflect the use of control to optimize both reward and duration (Shadmehr and Ahmed, 2020).

It has been similarly proposed that a core objective of cognitive control allocation is also the maximization of reward rate (Bogacz et al., 2006; Boureau et al., 2015; Lieder et al., 2018; Manohar et al., 2015; Shenhav et al., 2013). That is, that people select how much and what kinds of control to engage at a given time based on how control will maximize expected payoff (e.g., performance-based incentives like money or social capital) while minimizing the time it takes to achieve that payoff. Consistent with this proposal, studies have shown that people configure information processing (e.g., adjust their response thresholds) in ways that maximize reward rate (Balci et al., 2011; Simen et al., 2009; Starns and Ratcliff, 2010), and that they adjust this configuration over time based on local fluctuations in reward rate (Guitart-Masip et al., 2011; Otto and Daw, 2019).

We recently used a reward-rate optimization framework to make model-based predictions for how people coordinate *multiple* types of control (Leng et al., 2020). Participants performed a Stroop task that was self-paced, enabling them to dynamically adjust at least two forms of control: their overall drift rate (governing both how fast and accurate they are) and their response threshold (governing the extent to which they trade off speed for accuracy; Figure 1A). We varied the amount of money participants could gain with each correct response and the amount they could lose with each incorrect response. Participants could increase their response threshold to guarantee that every response was correct, but this came at the cost of completing fewer trials and therefore earning fewer rewards over the course of the. Increasing their drift rate can achieve higher reward rates, but is subject to effort costs, that we will return to later. The reward-rate optimal configuration across both drift and threshold would be to increase drift rate and decrease threshold for larger rewards, and increase thresholds for larger penalties (Figure 1B). Critically, we found that participants' DDM configuration matched the predictions of this optimal model (Figure 1C). These results provide evidence that participants' performance can align with the optimal joint configuration across multiple control parameters.

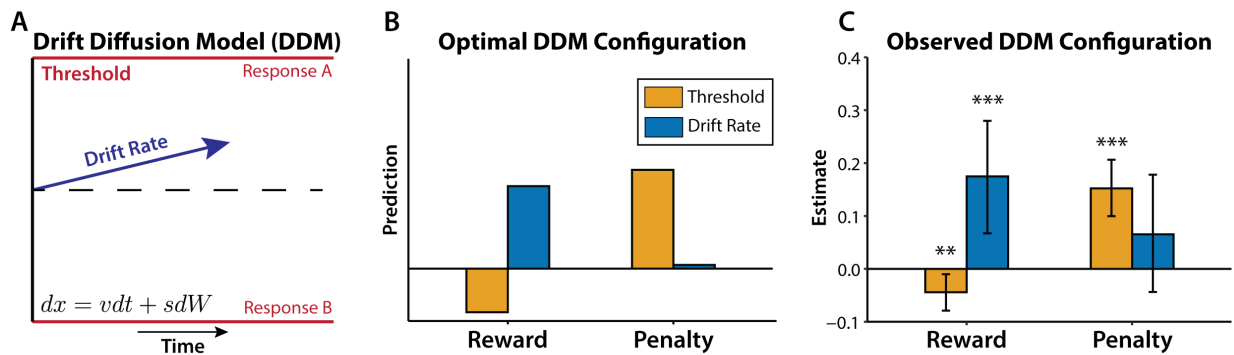

**Figure 1. Multivariate control configurations optimize reward rate. A)** In the drift diffusion model (DDM), the speed and accuracy of a decision are largely determined by the rate of evidence accumulation (drift rate; blue) and how much evidence the decision mechanism requires to make a choice (threshold; red). Evidence accumulates according to both the drift rate ($v$) and Gaussian diffusion noise ($s$). **B)** Leng and colleagues (2020) had participants perform a self-paced Stroop task, and examined how they adjusted their drift rate and threshold with varying levels of reward for correct responses and penalties for errors. A reward-rate optimal model predicted that higher rewards should bias their control configuration towards higher drift rates and lower thresholds, whereas larger penalties should bias these configurations towards higher thresholds and have little impact on drift **C)** DDM fits to task performance confirmed these predictions, demonstrating that participants adjusted their control configuration in a multivariate and reward-rate-optimal manner.

These studies validate the proposal that control allocation can be framed as decision-making over multidimensional configurations of control (i.e., combination of different control types engaged to different degrees) and that these decisions seek to optimize an objective function such as expected reward rate. While the DDM is useful for studying these configuration processes, providing a well-defined cognitive process model with criteria for good performance, similar optimality analyses have been performed in domains like working memory (Sims, 2015; Sims et al., 2012), demonstrating the generality of this approach. However, for all the algorithmic tools it provides, this decision-making framework also presents an entirely new set of challenges. Most notably, the many possible control configurations to choose from often means that there will be multiple equivalent solutions to this decision. Here, again, valuable insights can be gained from research on motor control, where these challenges and their potential solutions have been extensively explored.

# Inverse problems in motor and cognitive control

## Inverse problems in motor control

Some of the most influential computational modelling of motor planning was founded at the Central Labor Institute in Moscow in the early 20th century. This group formalized for the first time a fundamental problem for motor control: how does the motor system choose among the many similar actions that could be taken to achieve a goal (Bernstein, 1935; Whiting, 1983)? This problem is centered around the fact that motor control is inherently *ill-posed*, with more degrees of freedom in the body (e.g., joints) than in the task space, increasing the inherent challenge of selecting the best motor action among many equivalent options.

These motor redundancies can occur in several domains of motor planning (Kawato et al., 1990). At the task level, there may be many trajectories through the task space that achieve the same goals, such as the paths a hand could take on its way to picking up a cup (*Task Degeneracy*; Figure 2A). At the effector level, there are often more degrees of freedom in the skeletomotor system than in the task space, creating an 'inverse kinematics' problem for mapping from goals on to actions (*Effector Degeneracy*; Figure 2B). For example, there are many ways you could move your arm to trace a line with the tip of your finger. A related problem arises when there is redundancy across effectors, such as in agonist and antagonistic muscles (*Effector Antagonism*; Figure 2C). Due to their opponency, the same action can occur by trading off the contraction of one muscle against the relaxation of the other. These inverse problems have been a major challenge for theoretical motor control, and to the extent that a similar problem occurs in cognitive control, solutions from the motor domain may help guide our understanding of ill-posed cognitive control.

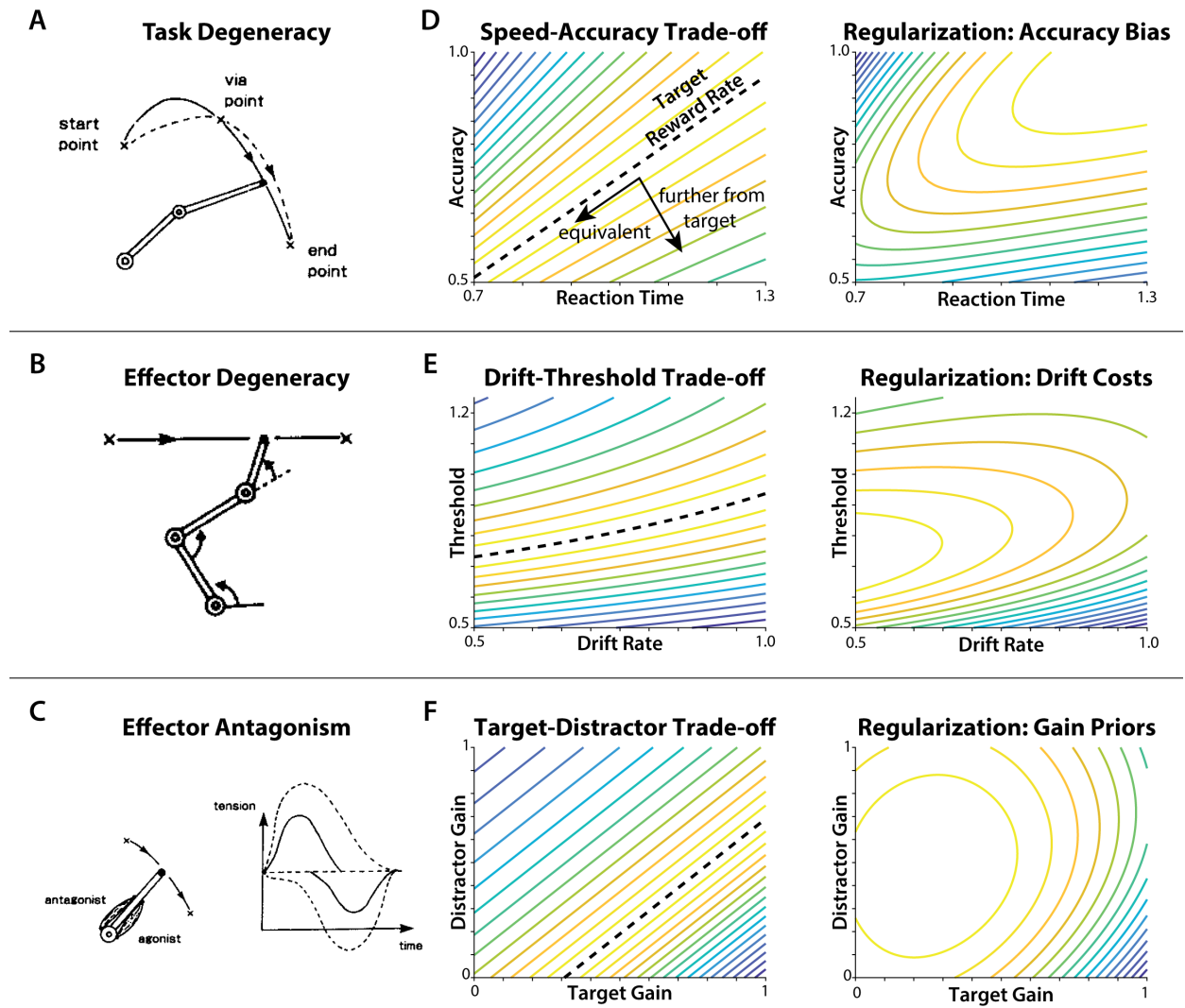

**Figure 2. Degeneracies in motor and cognitive control. A)** The fact that many trajectories can achieve the goal of moving from a start point to an endpoint during motor control results in *task degeneracy*. **B)** The fact that there are more degrees of freedom in the effectors (arm joints) than in the task (1D movement), and that there are many configurations that can produce the same movement, results in *effector degeneracy*. **C)**. The fact that some effectors have opposite influences over actions (e.g., agonist and antagonist muscles) results in *effector antagonism*. **D-F)** Analogous forms of degeneracy arise in relatively simple examples of cognitive control (**left** side of each panel)), such as when optimizing parameters of a DDM to achieve a target reward rate. Each of these forms of degeneracy can be solved in analogous way to motor control using different forms of regularization (**right** side of each panel). **D)** The target reward rate can be achieved with an infinite number of speed-accuracy trade-offs (points along dashed line), resulting in task degeneracy. A solution to this degeneracy is to include an additional preference for high accuracy, creating a globally optimal solution. **E)** Equivalent reward rates can also be achieved with various trade-offs between different model parameters being controlled (e.g., levels of drift rate and threshold), resulting in a form of 'effector' degeneracy. A solution to effector degeneracy is to place a cost on higher drift rates, biasing parameter configurations towards lower drift rates and creating a globally optimal solution. **F)** 'Effector' antagonism in cognitive control can result from opposing contributions of target gains (positive effect on drift rate) and distractor gains (negative effect on drift rate) on reward rate. A solution to effector antagonism is to set a prior on control gains, biasing these gains towards the prior configuration (e.g., high distractor sensitivity and low target sensitivity) and creating a globally optimal solution. Panels A-C adapted from (Kawato et al., 1990).

*Inverse problems in cognitive control: the algorithmic level*

Considering the massive degrees of freedom that exist in neural information processing systems, cognitive control is a prime candidate for inverse problems of its own. To illustrate this, we can return to the example of how people decide to allocate control across parameters of the drift diffusion model (Figure 2D). As reviewed above, participants can separately control individual parameters of evidence accumulation, specifically drift rate (Bond et al., 2021; Ritz and Shenhav, 2021), threshold (Cavanagh and Frank, 2014; Fischer et al., 2018), and accumulation noise (Mukherjee et al., 2021; Nakajima et al., 2019; Ritz et al., 2020). This test case of finding a reward-rate optimal configuration of DDM parameters faces the same set of challenges as those outlined above from motor control.

First, just as there are many hand trajectories that can produce a desired outcome, there are also many ways to produce good decision-making performance (Figure 2D). Different combinations of accuracy (numerator) and reaction time (denominator) can trade-off to produce the same reward rate. This creates an equivalence in the task space between different performance outcomes with regards to the goals of the system.

Second, just as there are more degrees of freedom in the arm than in many motor tasks, there is more flexibility in information processing than in many cognitive tasks. For example, the same patterns of behavior (and therefore expected reward rates) can result from different configurations of DDM parameters (Bogacz et al., 2006); Figure 2E). From a model-fitting perspective, this forces researchers to limit the parameters they attempt to infer from behavior, fixing at least one parameter value (often accumulation noise), while estimating the others (Bogacz et al., 2006; Ratcliff and Rouder, 1998). This degeneracy similarly limits a person's ability to perform the "mental model-fitting" required to optimize across all these control configurations when deciding how to allocate control. These difficulties are exacerbated in more biologically plausible models of evidence accumulation like the leaky competing accumulator (Usher and McClelland, 2001), which introduce additional parameters (e.g., related to memory decay and levels of inhibition across competing response units), resulting in even greater parameter degeneracy (Miletić et al., 2017). A similar trade-off exists in the classic debate between early and late attentional selection, namely whether attention operates closer to sensation or closer to response selection (Driver, 2001). Given that attention appears to operate at multiple processing stages (Lavie, 1995), degeneracies will arise if early or late attentional control will similarly influence task performance.

Third, just as there is antagonism across motor effectors, there is also antagonism across cognitive processes. That is, even when the algorithmic goal is clear, there are degenerate control signals that can achieve this goal. For instance, in typical interference-based paradigms (e.g., flanker or Stroop), participants must respond to one element of a stimulus while ignoring information that is irrelevant and/or distracting. To increase the overall rate of accumulation of goal-related information, the person can engage two different forms of attentional control: enhance targets or suppress distractors. Utilizing either of these strategies will improve performance, meaning that cognitive controller could trade off enhancing targets or suppressing distractors to reach the same level of performance (Figure 2F). Recent work has shown that target and distractor processing can be controlled independently in conflict tasks (Adkins and Lee, 2021; Evans and Servant, 2020; Ritz and Shenhav, 2021), creating an ill-posed problem of coordinating across these strategies.

*Inverse problems in cognitive control: the implementational level*

It is difficult to optimally configure decision-making, and this control faces several equivalent problems to those faced in motor control. In the case of algorithmic cognitive models, we find that parameter degeneracy (e.g., DDM) and process degeneracy (e.g., target-distractor trade-off) make it difficult to optimally configure information processing. However, problems at this level of analysis reflect the best-case scenario, as these cognitive models are themselves often intended to be lower-dimensional representations of the underlying neural processes (Bogacz, 2007). At the implementational level, cognitive control occurs over the complex neural instantiation of these algorithms, further exacerbating the ill-posed nature of the control problem.

One domain in which there can be redundancy in neural control is at the stage of processing at which control is applied, mirroring debates about early and late attentional selection highlighted above. Previous work has suggested that control can influence 'early' sensory processing (Adam and Serences, 2021; Egner and Hirsch, 2005) and 'late' processing in PFC (Mante et al., 2013; Stokes et al., 2013). To the extent that interventions along processing pathways have a similar influence on performance for a given task, there is a dilemma for where to allocate control.

The difficulty in deciding 'where' to allocate control is magnified as the control targets move from macro-scale processing pathways to local configurations of neural populations. For example, a controller could need to configure a small neural network to produce a specific spiking profile in response to inputs. Confounding this goal, it has been shown that a broad range of cellular and synaptic parameters produce very similar neuron- and network-level dynamics at the scale of only a few units (Alonso and Marder, 2019; Goaillard and Marder, 2021; Marder and Goaillard, 2006; Prinz et al., 2004). For example, very different configuration of sodium and potassium conductance can produce very similar bursting profiles (Golowasch et al., 2002), analogously to the redundancy of antagonistic muscles. These findings demonstrates that even simple neural networks face an ill-posed configuration problem, highlighting additional challenges to the biological implementation of cognitive control. Despite this degeneracy, research on brain-computer interfaces has shown that animals can exert fine-grained control over neural populations. Animals are capable of evoking arbitrary activity patterns to maximize reward (Athalye et al., 2019), even at the level of controlling single neurons (Prsa et al., 2017).

Across these different scales of implementation, the optimization of neural systems faces a core set of inverse problems. There are many macro-scale configurations that map similarly onto task goals, and there are many micro-scale configurations that map similarly on to local dynamics. This problem is closely related to the long-debated issue of *multiple realizability* in philosophy of science which, in its applications to neuroscience, has explored the lack of one-to-one mapping between neural and mental phenomena (e.g., whether pain is identical to 'C fiber' activity; (Putnam, 1967)). The lack of one-to-one mappings between structure and function poses not only an inferential problem to scientists and philosophers, but also an optimization problem to a brain's control system.

## The problem with inversion

As we've outlined above, the core difficulty in planning cognitive control comes from situations in which the brain needs to map a higher-dimensional control configuration on to a lower-dimensional task space, particularly when there is redundancy in this mapping (Figure 3). This class of problems has been carefully explored in applied mathematics (Calvetti and Somersalo, 2018; Engl et al., 1996; Evans and Stark, 2002; Willcox et al., 2021), and this field has developed helpful formalisms and solutions to the problems faced by the brain. We can first consider the *forward problem*, where a brain forecasts what would happen if it adopted a specific control configuration. For example, the controller may predict how performance will change if it raises its decision threshold. This problem generally has a unique solution, as a specific configuration will usually produce a specific result even if there is redundancy. Furthermore, projecting from a higher-dimensional configuration to a lower-dimensional outcome will compress the output, resulting in a stable solution.

However, the goal in optimization is to solve the *inverse problem*, in this case inferring which control configurations will produce a desired task state. As discussed earlier, this problem is generally ill-posed (Hadamard, 1902) because there are multiple redundant solutions for implementing cognitive control. Another reason this is an ill-posed problem is that this projects a lower-dimensional outcome into a higher-dimensional configuration (Calvetti and Somersalo, 2018; Engl et al., 1996). For example, the controller may optimize reward rate, but to do so must configure many potential neural targets. Since outcomes are noisy (e.g., noisy estimates of values due to sampling error or an imperfect forecasting), projection into a higher dimensional control space will amplify this noise. In this regime, small changes in values or goals can produce dramatically different control configurations, leading to an unstable optimization process. Without compensatory measures, these features of ill-posed cognitive control would impede the brain's ability to effectively achieve goals.

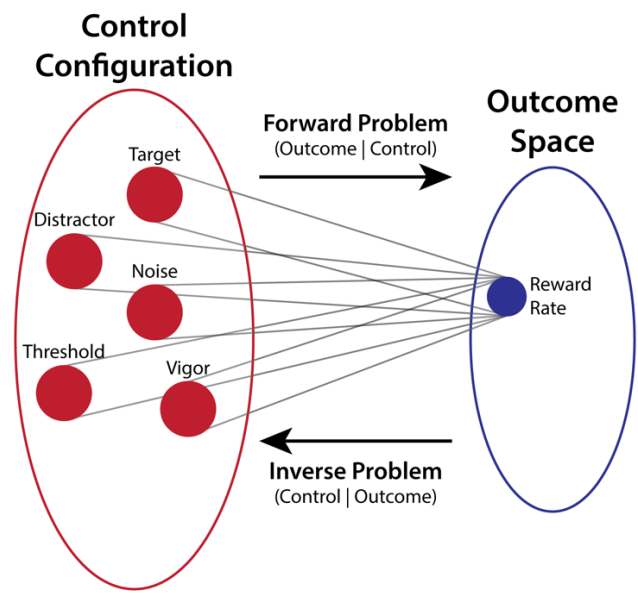

**Figure 3. Forward and inverse problems in cognitive control.** The forward problem in cognitive control entails predicting how a control configuration (**left**) would lead to a task state (**right**). This problem is stable

because it maps from a high-dimensional control space into a lower-dimensional task space. Specification of cognitive control requires solving the inverse problem, inferring the optimal control configuration to achieve a goal. This problem is unstable because it (redundantly) maps from a lower-dimensional task space into a higher-dimensional control space. Schematic adapted from (Krakauer et al., 2017).

This fundamental challenge of inferring the actions that will achieve goals has long been a central one within research on computational motor control (McNamee and Wolpert, 2019). Thankfully, these inverse problems can be made tractable though well-established modifications to the optimization process (Engl et al., 1996; Tikhonov, 1963). Motor theorists have leveraged these solutions to help explain action planning, and in doing so providing insight into the nature of effort costs.

# Solving the inverse problem

## *Motor solutions to the inverse problem*

A major innovation in theoretical motor control was to reframe the motor control problem as an optimization problem. Under this perspective, actions optimize an objective function over the duration of the motor action (similarly to the reward rate used for decision optimization). For scientists who took this approach, a primary focus was to understand people's objective functions, and in particular the costs that constrain people's actions. Researchers proposed that people place a cost on jerky movements (Flash and Hogan, 1985), muscle force (Chow and Jacobson, 1971; Nelson, 1983; Uno et al., 1989), or action-dependent noise (Harris and Wolpert, 1998), and therefore try to minimize one or more of these while pursuing their goals. A core difference between these accounts was whether costs depended on movement trajectories (Flash and Hogan, 1985) or muscle force (Uno et al., 1989), with the latter better explaining bodily constraints on actions (e.g., due to range of movement).

It now appears that actions are constrained by a muscle-force-dependent cost (Diedrichsen et al., 2010; Morel et al., 2017; O'Sullivan et al., 2009; Uno et al., 1989), and likely also endpoint noise (Harris and Wolpert, 1998; O'Sullivan et al., 2009; Todorov, 2005)). However, it remains unclear whether these effort costs are due to physiological factors like metabolism, or whether they reflect a more general property of the decision process. While metabolism would be an obvious candidate for these effort costs, researchers have found that subjective effort appraisals are largely uncorrelated with information being signaled by bodily afferents (Marcora, 2009). Furthermore, whereas metabolic demands should increase linearly with muscle force, effort costs are better accounted for by a quadratic relationship (Diedrichsen et al., 2010; Shadmehr and Ahmed, 2020).

These discrepancies suggest that motor effort may not depend solely on energy expenditure, but also on properties of the optimization process (e.g., related to the anticipated control investment). A promising explanation for these effort costs may arise from the solution to motor control's ill-posed inverse problem. A central method for solving ill-posed problems is to constrain the solution space through regularization (i.e., placing costs on higher intensities of motor control), a role that motor control theorists have proposed for effort costs (Jordan, 1989; Kawato et al., 1990). For example, across all motor plans that would produce equivalent performance outcomes, there is only one solution that also expends the least effort. From this perspective, motor effort enables better planning by creating global solutions to degenerate planning problems.

*Regularization as a solution to ill-posed cognitive control selection*

Much like motor control, cognitive control must also solve a degenerate inverse problem. Like motor control, cognitive control is subjectively costly (Shenhav et al., 2017). For example, participants will forego money (Westbrook et al., 2013) and even accept pain (Vogel et al., 2020) to avoid more cognitively demanding tasks. If physical effort regularizes degenerate motor planning, then it is plausible that cognitive effort similarly regularizes degenerate cognitive planning. Re-casting physical and mental effort as a regularization cost brings these domains in line with a wide range of related psychological phenomena. For example, inferring depth from visual inputs is also an ill-posed problem, and this inference has been argued to depend on regularization (Bertero et al., 1988; Poggio et al., 1985b, 1985a).

Recent proposals have drawn connections between cognitive effort and regularization under a variety of theoretical motivations. For instance, it has been proposed that cognitive effort enhances multi-task learning (Kool and Botvinick, 2018; Musslick et al., 2020), where effort costs regularize towards task-general policies ('habits') that enable better transfer learning. It has been also been proposed, based on principles of efficient coding (Zénon et al., 2019), that effort costs enable compressed and more metabolically efficient stimulus-action representations. Finally, effort costs have been motivated from the perspective of model-based control (Piray and Daw, 2021), where regularization towards a default policy allows for more efficient long-range planning. These accounts offer different perspectives on the benefits of regularized control, complementing motor control's emphasis on solving ill-posed inverse problems.

Regularization in inverse problems has a normative Bayesian interpretation, in which constraints come from prior knowledge about the solution space (Calvetti and Somersalo, 2018). This Bayesian perspective has been influential in modelling ill-posed problems like inferring knowledge from limited exemplars (Tenenbaum et al., 2011, 2006) and planning sequential actions (Botvinick and Toussaint, 2012; Friston et al., 2012; Solway and Botvinick, 2012). Regularization and Bayesian inference have been a productive approach for understanding how people solve ill-posed problems in cognition and action. Within the Bayesian frameworks, effort costs can be re-cast in terms of shrinkage towards a prior, providing further insight into how a regularization perspective could inform cognitive control. If there are priors on cognitive or neural configurations, such as automatic processes like habits, then regularized control would penalize deviations from those defaults.

A Bayesian perspective on the relationship between automaticity and control costs makes an interesting and counterintuitive prediction: when people's priors are to exert high levels of control, they will find it difficult to relax their control intensity. Research on control learning supports these predictions. A large body of work has found that participants learn to exert more control when they expect a task to be difficult (Bugg and Chanani, 2011; Jiang et al., 2015; Logan and Zbrodoff, 1979; Yu et al., 2009), or when stimuli are associated with conflict (Bugg and Crump, 2012; Bugg and Hutchison, 2013). This results in an allocation of excessive and maladaptive levels of control when a trial turns out to be easy (Logan and Zbrodoff, 1979). A recent experiment by Bustamante and colleagues (2021) extended these findings by showing how biases in control exertion can emerge through feature-specific reward learning. Participants performed a color-word Stroop task where they could choose to either name the color (more control-demanding) or read the word (less control-demanding). They learned that certain stimulus features would yield greater reward for color-naming and others for word-reading. Critically, during a subsequent transfer phase, participants had trouble learning to adaptively *disengage* control when faced with a combination of stimulus features that had each previously predicted greater reward for greater effort. That is, they had learned to *over-exert* control. However, further work is needed to understand whether this over-exertion is due to effort mobilization, or control priors that make color-naming less effortful (Yu et al., 2009).

This work highlights connections between control theory and forms of *reinforcement learning* that have been well-characterized within the cognitive sciences, whereby an agent is presumed to select actions (or sequences of actions) that maximize their expected long-term reward (Collins, 2019; Neftci and Averbeck, 2019; Sutton and Barto, 2018). Indeed, the parallels between these two modeling frameworks are rich, most notably in that both seek to optimize goal-directed behavior by optimizing the Bellman equation (a formula for estimating an action's expected future payoff; (Anderson and Moore, 2007; Kalman, 1960)). Ways in which these traditions often differ is that LQR emphasizes prospective model-based planning of a feedback policy over a continuous state space, whereas reinforcement learning usually focuses on gradually learning an action policy over a discrete state space (Recht, 2018). Reinforcement learning could speculatively intersect with cognitive control by learning the control priors highlighted above (complementing use-based automaticity (Miller et al., 2019) and evolutionary priors (Cisek, 2019; Zador, 2019)), or could be involved in learning higher-level control policies (e.g., learning a sequence of subgoals (Frank and Badre, 2012)).

# Algorithms for motor and cognitive control

Motor and cognitive control appear to solve similar problems (action-outcome inversion), and plausibly through similar computational principles (regularized optimization). The next logical step is to ask whether cognitive control has developed similar algorithmic solutions to this inversion as the motor control system. A longstanding gold-standard algorithm for modelling motor actions is the Linear Quadratic Regulator (LQR), which plays a central role in the Optimal Feedback Control theory of motor planning (Haar and Donchin, 2020; Shadmehr and Krakauer, 2008; Todorov and Jordan, 2002). Building off the success of Optimal Feedback Control in the motor domain, this algorithm provides a promising candidate for understanding the planning and execution of *cognitive* actions.

LQR can provide the optimal solution to sequential control problems when two specific criteria are met. First, the system under control must have linear dynamics, such as a cruise controller that adjusts the speed of a car. Second, the control process must be optimizing a quadratic objective function. This usually involves minimizing both the squared goal error (e.g., the squared deviation from desired speed) and the squared control intensity (e.g., the squared motor torque). Under these conditions, LQR provides an analytic (i.e., closed-form) solution to the optimal policy[3], avoiding the curse of dimensionality (Van Rooij, 2008). LQR is equivalent to the Kalman filtering method for optimal inference (Kalman and Bucy, 1961; Todorov, 2008), and the Linear Quadratic *Gaussian* algorithm combines inference and control for computationally tractable optimal behavior under state uncertainty (Todorov, 2005; Yeo et al., 2016).

In the domain of motor control, this model empirically captures participants' motor trajectories (Stevenson et al., 2009; Todorov and Jordan, 2002; Yeo et al., 2016), particularly in the case where there are mid-trajectory perturbations to goals or effectors (Diedrichsen, 2007; Knill et al., 2011; Liu and Todorov, 2007; Nashed et al., 2012; Takei et al., 2021). A striking example of the power of this model to capture behavior was observed in an experiment on motor coordination (Diedrichsen, 2007). Participants performed a reaching task in which the goal either depended on both arms (e.g., rowing), or where each arm had a separate goal (e.g., juggling). During the reach, the experimenters perturbed one of the arms, and found that participants compensated with both arms only when they were both involved in the same goal. In LQR, this goal-dependent coordination arises due to the algorithm's model-based feedback control, with squared effort costs favoring distributing the work across goal-relevant effectors. Accordingly, this study found that LQR simulations accurately captured participants' reach trajectories. Furthermore, participants' behavior also confirmed a key prediction of LQR, namely that noise correlations between arms will be task-specific, constraining control to the goal-relevant dimensions of the task manifold (the 'minimal intervention principle'; (Todorov and Jordan, 2002)).

A starting point for developing algorithmic links between cognitive and motor control is to consider whether cognitive control is a problem that is well-suited for LQR. The first prediction from LQR is that the dynamics between cognitive states are approximately linear. One measure of these dynamics comes from task switching, in which participants switch between multiple stimulus-response rules ('task sets'; (Monsell, 2003)). Researchers have found that these transitions between task sets are well-captured by linear dynamics (Musslick et al., 2019a; Musslick and Cohen, 2021; Steyvers et al., 2019). For example, when participants are given a variable amount of time to prepare for a transition between two tasks (e.g., responding based on letters vs digits), the stereotypical switch cost of slower responding after a task switch compared to a task repetition decreases with greater preparation time (Rogers and Monsell, 1995). A simple re-analysis of this pattern shows that switch costs can be well-captured by a linear dynamical model (Figure 4A). Whereas switching to the 'letter' or 'digit' task had different initial and asymptotic performance costs, they appear to exhibit a similar rate of change.

Linear dynamics have also been observed in attentional adjustments that occur *within* a trial of a given task. For instance, recent work has shown that performance on an Eriksen flanker task can be accounted for by a DDM variant in which initially-broad attention narrows within a trial to primarily focus only the central target, resulting in a shift from the drift rate being initially dominated by the flankers to being primarily dominated by the target (Servant et al., 2014; Weichart et al., 2020; White et al., 2011). Using the dot motion task described earlier, we recently showed that these within-trial dynamics can be further teased apart into target-enhancing and distractor-suppressing elements of feature-based attention, each with its own independent dynamics (Ritz and Shenhav, 2021). These dynamics were well-captured by an accumulation model that regulated feature gains with a linear feedback control law (Figure 4B).

---

[3] The analytic solutions to these algorithms rely on ordinary least squares solutions for optimizing quadratic loss functions and Gaussian identities describing how quadratic loss functions change under linear dynamics. For in-depth mathematical treatments, see (Anderson and Moore, 2007; Recht, 2018; Shadmehr and Krakauer, 2008).

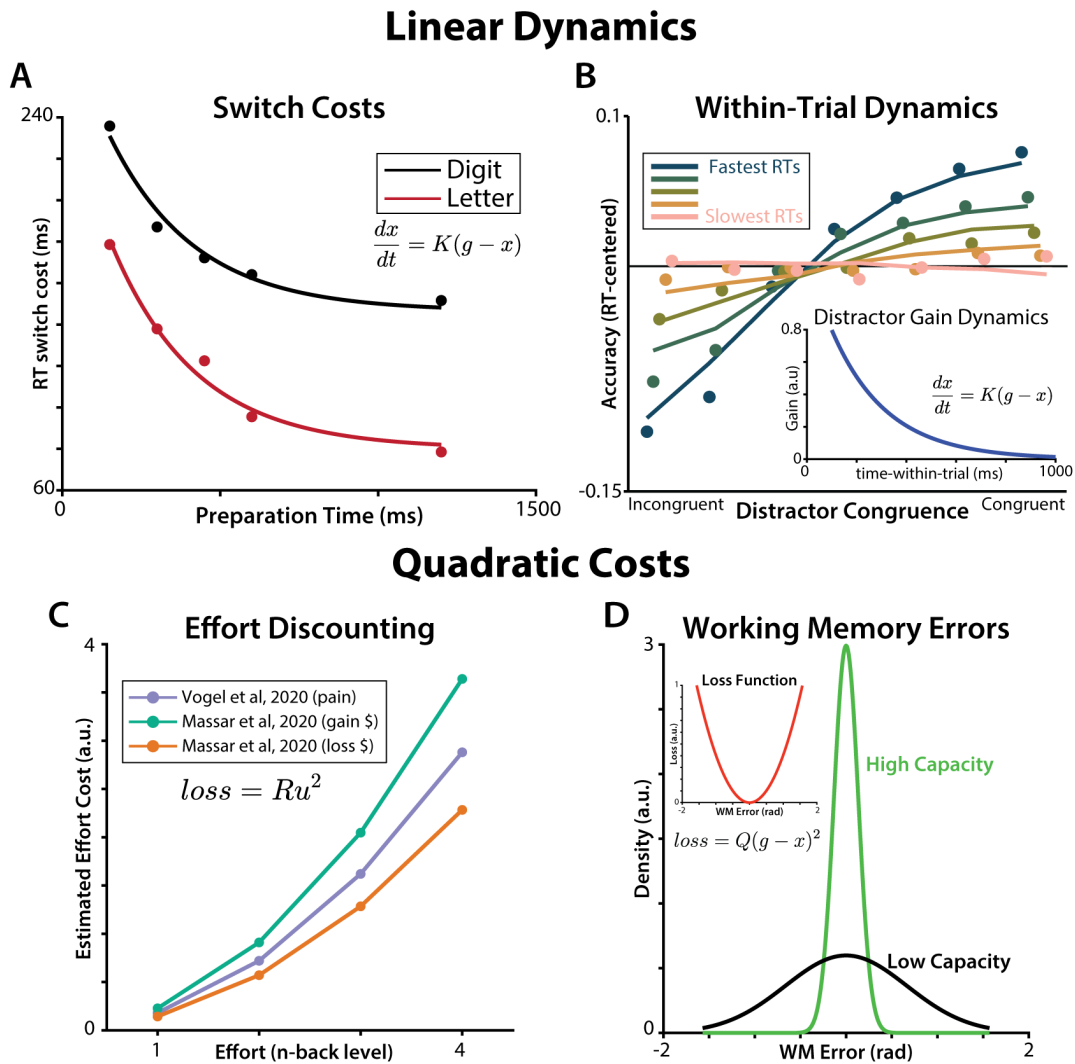

**Figure 4**. **Linear-Quadratic properties of cognitive control. A-B)** There is evidence of linear cognitive control reconfiguration dynamics both between trials and within a given trial. **(A)** In task-switching experiments, participants' switch costs (slower and less accurate performance when performing a different task than on the previous trial) exponentially decay with longer preparation time (time between the end of one trial and the start of the next), consistent with linear dynamics. Lines show a maximum likelihood fit to data from Rogers and Monsell (1995) in which participants switched between letter and digit tasks at predictable intervals. We estimated a shared decay rate (K) across tasks, with separate initial conditions and asymptote fit to average switch costs in each task. **(B)** In a response conflict task, participants were less sensitive to distractor conflict (parametrically varying stimulus-response congruence) at later response times (Ritz and Shenhav, 2021). This experiment modelled participants' distractor sensitivity dynamics as exponentially decaying over time within each trial (inset), consistent with linear dynamics (Weichart et al., 2020; White et al., 2011). **C-D)** Quadratic cost functions are evident in studies of effort discounting and working memory. **(C)** In effort-discounting tasks,

participants' subjective cost of n-back tasks quadratically increases with their working memory load. Estimated cost functions are plotted from (Massar et al., 2020; Vogel et al., 2020). **(D)** Errors on working memory tasks are approximately Gaussian, consistent with a quadratic loss function on accuracy (Sims et al., 2012).

A second prediction from LQR is that cognitive effort costs are quadratic. There are two lines of evidence that support this prediction. One line of evidence comes from studies of *cognitive effort discounting*, which examine how people explicitly trade off different amounts reward (e.g., money) against different levels of cognitive effort (e.g., n-back load). These studies quantify the extent to which different levels of effort are treated as a cost when making those decisions (i.e., how much reward is discounted by this effort), and many of them find that quadratic effort discounting captures choice the best among their tested models[4] (Figure 4C; (Białaszek et al., 2017; Massar et al., 2020; Petitet et al., 2021; Soutschek et al., 2014; Vogel et al., 2020); though see also (Chong et al., 2017; Hess et al., 2021; McGuigan et al., 2019). A second line of evidence supporting quadratic costs is found in tasks that require participants to hold a stimulus in working memory (e.g., a Gabor patch of a given orientation) and then reproduce that stimulus after a delay period. Errors on this task tend to be approximately Gaussian (Bays and Husain, 2008; Ma et al., 2014; Sprague et al., 2016; van den Berg et al., 2012; Wilken and Ma, 2004), consistent with the predictions of ideal observer models that incorporate quadratic loss function (Sims, 2015; Sims et al., 2012); Figure 4D).

Recent work has begun to make explicit links between LQR and the neural implementation of cognitive control. Most notably, Bassett and colleagues have used LQR to model the large-scale control of brain networks (Tang and Bassett, 2018)). This approach uses LQR modelling of whole-brain network dynamics to understand the ability of sub-networks to reconfigure macro-scale brain states (Betzel et al., 2016; Braun et al., 2021; Gu et al., 2021, 2015), see also (Yan et al., 2017). For instance, in an fMRI experiment using the n-back task, Braun and colleagues (2021) found that their LQR model inferred that the brain requires more control to maintain a stable 2-back state than a 0-back state, as well as more control to transition from a 0-back state into a 2-back state than vice versa. Interestingly, individual differences in these model-derived estimates of stability and flexibility were associated with differences in dopamine genotype, dopaminergic receptor blockade, and schizophrenia diagnosis (Braun et al., 2021). An LQR modelling approach has been similarly used to model dynamics in directly-recorded neural activity to understand how local connectivity influences control demands (Athalye et al., 2021; Stiso et al., 2019), with accompanying theories of how these configuration processes are learned through reinforcement learning (Athalye et al., 2019).

# Conclusions and future directions

The second half of the twentieth century saw a wave of progress on mathematical models for optimal control problems in applied mathematics. A second wave of computational motor control followed closely, combining rigorous measurement of motor actions with normative models from this new optimal control theory (Chow and Jacobson, 1971; Flash and Hogan, 1985; Nelson, 1983; Todorov and Jordan, 2002; Uno et al., 1989). Recently, a third wave of cognitive control research has extended optimal control principles to goal-directed cognition (Bogacz et al., 2006; Lieder et al., 2018; Musslick and Cohen, 2021; Piray and Daw, 2021; Shenhav et al., 2017, 2013; Tang and Bassett, 2018; Yu et al., 2009). This work tries to formalize the principles that tie these different frameworks together, highlighting how cognitive control can learn from decades of computational motor control research. These principles have the potential to inform the theoretical development and focused empirical investigation into the architecture of goal-directed cognition. As behavioral tasks, statistical techniques, and neuroimaging methods improve our measurements of how the brain configures information processing, theoretical constraints will be essential for asking the right questions.

---

[4] A concern about effort discounting is that it ought to be estimated based on cognitive demands rather than task demands. Notably, participants consistently show quadratic effort discounting in the n-back task, one domain where there is at least a well-characterized linear relationship between these levels of task load and PFC activity (Braver et al., 1997).

One insight from casting cognitive control as regularized optimization is that the sources of the control costs that can enable 'failures' of control are not necessarily due to cognitive *limitations* (e.g., limited capacity to engage multiple control signals). Instead, these costs can arise due to the *flexibility* of cognition, enabling a complex brain to optimize over degenerate control actions. Under this framework, effort helps *solve* the decision problem of how to configure control. One productive application of this perspective may be to help shed light on why people differ in how they configure these multivariate signals, for instance prioritizing some forms of control over others. A regularization perspective would emphasize understanding different people's priors (such as judgements of ability, (Bandura, 1977; Shenhav et al., 2021)) and configural redundancy when accounting for people's mental effort costs.

There are several important avenues for building further on the promising theoretical and empirical foundations that have been recently established in the study of multivariate control optimization. For instance, it will be important to understand how effort's role in solving the inverse problem trades off against other proposed benefits like generalization (Kool and Botvinick, 2018; Musslick et al., 2020) and efficiency (Zénon et al., 2019). It will also be important to develop finer-grained connections between computational theories of regularized cognitive control and the algorithmic and implementational theories of how the brain performs control optimization and execution. For instance, to what extent can specific regularized control algorithms such as LQR explain the dynamics of cognitive control optimization and deployment? How does the cognitive control system integrate across multiple monitored signals of goal progress and achievement (Haar and Donchin, 2020), including different forms of errors and conflict (Ebitz and Platt, 2015; Shen et al., 2015)? While LQR modelling has been a powerful approach for understanding the role of neural connectivity in goal-driven brain dynamics, more work is needed to bridge these findings to cognitive models of control optimization and specification.

In addition to understanding the computational goals of cognitive control optimization, it will be equally important to understand how biological control algorithms deviate from optimality. A substantial body of research has characterized apparent deviations from optimality during judgment and decision-making in the form of heuristics and biases (Kahneman, 2003; Tversky and Kahneman, 1974). Such seemingly irrational behaviors have been accounted for within decision frameworks by formalizing the rational bounds on optimality (Bhui and Xiang, 2021; Gershman and Bhui, 2020; Lieder et al., 2014, 2012; Lieder and Griffiths, 2019; Parpart et al., 2018; Simon, 1955). The LQR algorithm may similarly reflect bounded optimality, as LQR is sub-optimal when its linear-quadratic assumptions are a poor match to a task. A cognitive control system that uses LQR could reflect a trade-off between better computational tractability and poorer worst-case performance. Future research should investigate how the heuristics, biases, and approximations that influence cognitive control can inform our models of control planning.

Progress on these questions will in turn require more precise estimates of the underlying control processes. The study of motor control has benefited immensely from high resolution measurements of motor effectors, for instance tracking hand position during reaching. Analogous measures of cognitive control are much more difficult to acquire, in part because they require inference from motor movements (e.g., response time) and/or patterns of activity within neural populations whose properties are still poorly understood and are typically measured with limited spatiotemporal resolution. Future experiments should combine computational modelling with spatiotemporally resolved neuroimaging to understand the implementation of different types of control. In addition to addressing core questions at the heart of multivariate control optimization, such methodological improvements will also help us better understand the heterogeneity of multivariate effort. For instance, an untested assumption implied by existing theoretical frameworks is that all forms of cognitive control will incur subjective costs in a similar fashion, for instance that higher levels of drift rate and higher levels of threshold will both be experienced as effortful (cf. (Shenhav et al., 2013)). While there is consistent evidence that enhancements to drift rate incur a cost, it remains less clear whether adjustments to response threshold incur a cost over and above the reductions to reward rate they can cause (cf. Leng et al., 2020). Further research is needed to examine this question and to explore both the magnitude and functional form of these cost functions across a wider array of control signals, especially with respect to deviations from participants' default configurations.

Our cognitive control is extremely complex, flexible, primarily operates over latent processes like decision making, all features that make studying cognitive control a challenge. Thankfully, we can gain better traction on this inference by drawing from the rich empirical and theoretical traditions in better-constrained fields like motor control (Broadbent, 1977). The normative principles of optimal control theory, which have proven so fruitful in motor control, can similarly help inform our theories and investigations into cognitive control. While our cognition will certainly diverge from these normative theories, these approaches can provide a core foundation for understanding how we control our thoughts and actions.

**Acknowledgements:** This work was supported by the Training Program for Interactionist Cognitive Neuroscience T32-MH115895 (X.L.), and grants R01MH124849 and NSF CAREER Award 2046111 (A.S.). Special thanks to Laura Bustamante, Romy Frömer, and the rest of the Shenhav Lab for helpful discussions on these topics.

Meeting of the Cognitive Science Society.

Ritz H, Shenhav A. 2021. Humans reconfigure target and distractor processing to address distinct task demands. *bioRxiv*. https://doi.org/10.1101/2021.09.08.459546

Rogers RD, Monsell S. 1995. Costs of a predictable switch between simple cognitive tasks. *J Exp Psychol Gen* **124**:207.

Schroeder U, Kuehler A, Haslinger B, Erhard P, Fogel W, Tronnier VM, Lange KW, Boecker H, Ceballos-Baumann AO. 2002. Subthalamic nucleus stimulation affects striato-anterior cingulate cortex circuit in a response conflict task: a PET study. *Brain* **125**:1995–2004. https://doi.org/10.1093/brain/awf199

Servant M, Montagnini A, Burle B. 2014. Conflict tasks and the diffusion framework: Insight in model constraints based on psychological laws. *Cogn Psychol* **72**:162–195. https://doi.org/10.1016/j.cogpsych.2014.03.002

Shadmehr R, Ahmed AA. 2020. Vigor: Neuroeconomics of movement control. MIT Press.

Shadmehr R, Krakauer JW. 2008. A computational neuroanatomy for motor control. *Exp Brain Res* **185**:359–381. https://doi.org/10.1007/s00221-008-1280-5

Shadmehr R, Orban de Xivry JJ, Xu-Wilson M, Shih T-Y. 2010. Temporal discounting of reward and the cost of time in motor control. *J Neurosci* **30**:10507–10516. https://doi.org/10.1523/JNEUROSCI.1343-10.2010

Shen C, Ardid S, Kaping D, Westendorff S, Everling S, Womelsdorf T. 2015. Anterior Cingulate Cortex Cells Identify Process-Specific Errors of Attentional Control Prior to Transient Prefrontal-Cingulate Inhibition. *Cereb Cortex* **25**:2213–2228. https://doi.org/10.1093/cercor/bhu028

Shenhav A, Botvinick MM, Cohen JD. 2013. The expected value of control: an integrative theory of anterior cingulate cortex function. *Neuron* **79**:217–240. https://doi.org/10.1016/j.neuron.2013.07.007

Shenhav A, Musslick S, Lieder F, Kool W, Griffiths TL, Cohen JD, Botvinick MM. 2017. Toward a rational and mechanistic account of mental effort. *Annu Rev Neurosci* **40**:99–124.

Shenhav A, Prater Fahey M, Grahek I. 2021. Decomposing the Motivation to Exert Mental Effort. *Curr Dir Psychol Sci* **30**:307–314. https://doi.org/10.1177/09637214211009510

Shiffrin RM, Schneider W. 1977. Controlled and automatic human information processing: II. Perceptual learning, automatic attending and a general theory. *Psychol Rev* **84**:127.

Simen P, Contreras D, Buck C, Hu P, Holmes P, Cohen JD. 2009. Reward rate optimization in two-alternative decision making: empirical tests of theoretical predictions. *J Exp Psychol Hum Percept Perform* **35**:1865–1897. https://doi.org/10.1037/a0016926

Simon HA. 1955. A Behavioral Model of Rational Choice. *The Quarterly Journal of Economics*. https://doi.org/10.2307/1884852

Sims CR. 2015. The cost of misremembering: Inferring the loss function in visual working memory. *J Vis* **15**. https://doi.org/10.1167/15.3.2

Sims CR, Jacobs RA, Knill DC. 2012. An ideal observer analysis of visual working memory. *Psychol Rev* **119**:807–830. https://doi.org/10.1037/a0029856

Solway A, Botvinick MM. 2012. Goal-directed decision making as probabilistic inference: a computational framework and potential neural correlates. *Psychol Rev* **119**:120–154. https://doi.org/10.1037/a0026435

Soutschek A, Stelzel C, Paschke L, Walter H, Schubert T. 2015. Dissociable effects of motivation and expectancy on conflict processing: an fMRI study. *J Cogn Neurosci*